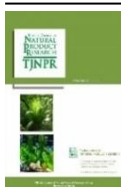
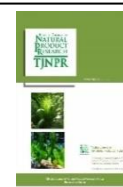

# Tropical Journal of Natural Product Research

Available online at https://www.tjnpr.org

*Original Research Article*

# Antibacterial and Antibiofilm Activities of a Traditional Herbal Formula against Respiratory Infection Causing Bacteria

Haitham Qaralleh[1]*, Muhamad O. Al-Limoun[2], Ali Khlaifat[3], Khaled M. Khleifat[2], Nafe Al-Tawarah[1], Khalid Y. Alsharafa[2], Hashem A. Abu-Harirah[4]

[1]*Department of Medical Laboratory Sciences, Mutah University, Mutah, Karak, Jordan*
[2]*Biology Department, Mutah University, Mutah, Karak, Jordan*
[3]*Department of Nursing, Faculty of Prince Aysha for Applied Health and Nursing, Al-Hussein Bin Talal University, Ma'an, Jordan*
[4]*Faculty of Allied medical Sciences- Zarqa University, Jordan*

ABSTRACT

The plants, *Althaea officinalis*, *Tilia cordata* and *Psidium guaja* have been used traditionally to treat respiratory infection symptoms. Flowers of *A. officinalis* and leaves of *T. cordata* and *P. guaja* have been used to treat cough, sore throat, catarrh, oral and pharyngeal mucosa irritation. Therefore, this study was designed to examine the antibacterial and antibiofilm effects of these plants individually as well as in combination, as a formula against respiratory infections causing pathogens. The tested pathogens were Extended Spectrum Beta-Lactamase producing *Escherichia coli* (ESBL), Beta-Lactamase producing *Escherichia coli* (BL), Beta-Lactamase producing *Klebsiella pneumoniae* (BL), Beta-Lactamase producing *Pseudomonas aeruginosa* (BL), *Enterobacter cloacae*, and Beta-Lactamase producing *Staphylococcus aureus* (BL). The tested plants were extracted using ethanol and then fractionated using different polarity solvents (hexane, ethyl acetate and water). Disc diffusion and microdilution (Minimum Inhibitory Concentration) methods were used to evaluate the antibacterial activity while the antibiofilm activity was tested using crystal violet assay. The results showed that *A. officinalis* and *T. cordata* extracts and fractions exhibited weak antibacterial activity having MIC values ranged from 6.25 to 12.5 mg/mL. *P. guaja* exhibited moderate antibacterial activity with MIC values ranged from 6.25 to 1.56 mg/mL. Combination between these plants extracts and fractions in equal proportion provides stronger antibacterial (with MIC values ranged from 6.25 to 0.8 mg/mL) and antibiofilm activities ($MBIC_{50}$ was 0.2 mg/mL). Therefore, this study provides a valuable scientific knowledge to support the use of plants in combination rather than individually.

## Introduction

Respiratory infections are considered as the most common infection worldwide. Globally, more than 50 million deaths are reported every year due to infections related to respiratory system.[1,2] The respiratory infections could be caused by several pathogenic agents including viruses and bacteria.[3,4] In case of bacterial infections, most common causative agents for pharyngitis and tonsillitis are Group A Beta Hemolytic Streptococci, *Corynebacterium diphtheria* and *Neisseria gonorrhoeae*. Epiglottitis and bronchitis are frequently caused by *Haemophilus influenzae* type b, *Corynebacterium diphtheria*, *Streptococcus pneumoniae* and *Mycoplasma pneumoniae*. Pneumonia is caused by such agents as *Streptococcus pneumoniae, Staphylococcus aureus, Streptococcus pyogenes, Haemophilus influenzae, Klebsiella pneumoniae, Escherichia coli, Pseudomonas aeruginosa* and *Mycobacterium tuberculosis*.[5] In fact, a steady increase in the frequency of antibiotics resistant bacteria implicated in respiratory infections is noteworthy. Due to this problem, in 2017, the World Health Organization issued a list of pathogens that required immediate action to develop new antibiotic therapy.[6] In addition to carbapenem-resistant *Acinetobacter baumannii* and carbapenem-resistant *Pseudomonas aeruginosa* issued in this list, carbapenem- and third-generation cephalosporin-resistant have been considered. They later include *Klebsiella pneumoniae*; methicillin resistant and vancomycin resistant *Staphylococcus aureus*; penicillin nonsusceptible *Streptococcus pneumoniae*; and ampicillin-resistant *Haemophilus influenzae*. Therefore, finding new effective antimicrobial agents to treat respiratory infections is highly required.

In Jordan, recent studies conducted in two districts, reported that the most common infections in elderly patients were urinary tract infection (53.75%), and respiratory tract infection (23.75%).[7,8] An increasing interest in plants traditional medicine has emerged. Traditional medicine is a fundamental source for treating diseases worldwide.[9,10] In developing countries, more than 70% of the population still rely on traditional medicine.[11] In Jordan, report showed that more than 65% of the populace think that traditional medicine is effective in curing diseases.[12] However, many reports had documented the traditional use of medicinal plants including traditional healers as key sources for these traditional knowledge.[10,13] Usually, individual or combined extracts of plant(s) are recommended to treat diseases. Frequently, combined formula of several traditional medicinal plants is used potently for medication with lower toxicity.[14]
In Jordan, the flowers from *A. officinalis*, leaves from *T. cordata* and *P. guaja* are recommended frequently by traditional healers to treat

*Corresponding author. E mail: haitham@mutah.edu.jo
Tel: 00962-797489248

527

respiratory infections. The traditional uses of these plants are broad and some of these uses are associated with the respiratory system infections.

*A. officinalis* (English name: Marshmallow Plant; Local name: Khetmeah) is a perennial species that belongs to Malvaceae family. It is native throughout Europe, Western Asia, and North Africa. Traditionally, decoction of the dried leaves and flowers is used to treat dry cough, oral and pharyngeal mucosa irritation and as expectorant.[15,16] Several secondary metabolites were isolated from *A. officinalis* including coumarins, scopoletin, hypolaetin-8-glucoside, Isoquercitrin, kaempferol, caffeic, p coumaric acid, ferulic acid, p-hydroxybenzoic acid, salicylic acid, p-hydroxyphenylacetic aicd and vanillic acid.[16]

*T. cordata* (English name: lime tree; local name: Zezafoon) is deciduous tree that belongs to Malvaceae family. Traditionally, leaves is prepared as a herbal tea to relief the symptoms of common cold, coughs and catarrh.[17] Lime tree possesses a diverse groups of secondary metabolites including caffeic, p-coumaric, chlorogenic acids, kaempferol, quercetin, myricetin, hyperoside, quercitrin, isoquercitrin, anethole, citral, citronellol, eugenol, limonene, menthone, nerol, a-pinene, terpineol, fenchone, a- and b-thujone, and farnesol.[18]

*P. guaja* (English name: Guajá; local name: Guava) is belonging to Myrtaceae family. It is a fruit tree that distributed all over the world. Decoction of leaves of *P. guaja* is used orally or as gargle to treat flu, sore throats and cough.[19–21] The leaves of *P. guaja* is rich in menthol, α-pinene, β-bisabolene, β-pinene, β-copanene, limonene, terpenyl acetate, isopropyl alcohol, caryophyllene, longicyclene, cineol, caryophyllene oxide, humulene, farnesene, selinene, curcumene and cardinene.[22]

Moreover, many pharmacological investigations have confirmed that the extracts and isolated compounds from *A. officinalis*, *T. cordata* and *P. guaja* possess broad biological activities in which this supports their broad and extensive traditional uses. Pharmacological investigations showed that *A. officinalis* possess anti-complement, anti-inflammatory, antitussive, antioxidant and hypoglycemic effects while *T. cordata* possess anti-fungal, anti-viral, anti-inflammatory and Antioxidant activities.[16,23] Gutiérrez *et al*.,[21] and Anand *et al*.,[24] reported that *P. guaja* are effective as antioxidant, hepatoprotection, anti-allergy, antimicrobial, antigenotoxic, antiplasmodial, cytotoxic, antispasmodic, cardioactive, anticough, antidiabetic, antiinflamatory and antinociceptive agent.

Therefore, this study was designed to examine the inhibitory effect of the flowers from *A. officinalis*, leaves from *T. cordata* and *P. guaja* individually as well as in combination as a formula against pathogens that caused respiratory infections. The ability of these plants extracts to inhibit *P. aeruginosa* and *E. coli* biofilms was also evaluated.

## Materials and Methods

*Plant materials and extraction*
The flowers from *A. officinalis*, leaves from *T. cordata* and leaves from *P. guaja* were purchased from herbal markets (Mutah, Alkarak, Jordan). Voucher specimens (No. M120, M121 and M122, respectively) were deposited in the department of medical laboratory sciences, Mutah University, Jordan. The plants parts were dried in shade for 10 days then they were cleaned and ground into fin powder. From the powder plant, 250 g was weighed into flask containing 600 mL of 95% ethanol and kept at 20-25°C for 24 h. The crude ethanol extracts for each plant were collected, filtrated using Millipore filter syringe (0.45 µm) and stored at 4°C.

*Liquid-liquid fractionation*
Liquid-liquid fractionation of the crude ethanol extracts for each plant were performed according to Bibi *et al*.[25] with some modification. Briefly, each extract (10 g) was suspended in 250 mL water and partitioned sequentially with two organic solvents (hexane and ethyl acetate, 200 mL, each) using separating funnel. Solvents of the ethanol extract and those of three fractions (hexane, ethyl acetate and water) were removed, filtrated and appropriately concentrated using rotary evaporator. Finally, all extracts and fractions (hexane, ethyl acetate and water) were stored as aliquots at 4°C.

*Preparation of test extracts*
Stock solutions were prepared by dissolving 100 mg from the test extracts and fractions (ethanol, hexane, ethyl acetate and water) separately in 1000 µL dimethyl sulfoxide (DMSO) and sequentially filtrated through 0.22 µm filter syringe. In addition, similar stock solutions were prepared for the herbal formula that composed of a mixture of extracts (ethanol, hexane, ethyl acetate and water) in equal proportion 1:1:1. All samples were stored at -4°C.

*Antibacterial activity of plant extracts*
Five Gram negative strains including Extended Spectrum Beta-Lactamase producing *Escherichia coli* (ESBL), Beta-Lactamase producing *Escherichia coli* (BL), Beta-Lactamase producing *Klebsiella pneumoniae* (BL), Beta-Lactamase producing *Pseudomonas aeruginosa* (BL) and *Enterobacter cloacae*, and one Gram positive strain; Beta-Lactamase producing *Staphylococcus aureus* (BL) were used in this study. These strains were provided by Al Bashir hospital (Amman, Jordan). In addition, two reference strains: *E. coli* ATCC 25922 and *P. aeruginosa* ATCC 10145 provided by Department of Biology, Mu'tah University, Al-Karak, Jordan were used.

The disc diffusion method was performed as previously described[26–28] with some modifications. Briefly, 24 h bacterial suspension adjusted to 0.5 McFarland's standard ($1.5 \times 10^8$ CFU/mL) was prepared. Then, Mueller-Hinton agar plates were seeded with 100 µL of the suspended bacteria and spreading was performed using sterile cotton swab. A sterile blank discs (6 mm in diameter) containing 0.5 mg of plant extracts, negative control (DMSO) or Ciprofloxacin (5 µg) were transferred into the inoculated plates. All the plates were incubated at 37°C for 24 h. After which, the inhibition zone formed was measured as millimeter diameter. Each test was performed in triplicate.

*Minimum inhibitory concentration (MIC) of plant extracts*
Microdilution method was used to estimate the Minimum inhibitory concentration (MIC) of each extract as previously described[29] with some modifications. A 96-well plate was used to prepare two-fold dilution of the tested extracts. The final concentrations of the extracts were 0.2, 0.4, 0.8, 1.56, 3.13, 6.25, 12.5 and 25.0 mg/mL. Then, 10 µL of bacterial suspension adjusted to 0.5 McFarland's standard ($1.5 \times 10^8$ CFU/mL) was inoculated into each well. The same test was carried out with DMSO as a control. The inoculated plates were incubated at 37°C for 24 h. Each test was performed in triplicate. The lowest concentration of the tested extracts needed to inhibit the visible growth of the tested microbes after 24 h was considered as the MIC values.

*Anti-Biofilm activity and minimum biofilm inhibitory concentration (MBIC)*
The anti-biofilm activity against *E. coli* (ESBL), and *P. aeruginosa* (BL), was evaluated using crystal violet assay.[30] Sub-MIC concentrations equal to 0.2, 0.4, 0.8 mg/mL were prepared using 96-well plate. Then 10 µL of bacterial suspension containing $1.5 \times 10^8$ CFU/mL (0.5 McFarland's standard) was inoculated into each well. The same test was carried out with DMSO as a control. Each test was performed in triplicate. After 24 h incubation at 37°C, the growth medium was discarded, and the plates were washed with distilled water and stained with 200 µL of 0.4% crystal violet. After 20 min, the stain was removed, and the excess stain was rinsed off with tap water (three times, 200 µL, each) before adding 200 µL of 95% (v/v) ethanol to solubilize the crystal violet. From the dissolved crystal violet into ethanol, 150 µL from each well were transferred to new 96 well plate for spectrophotometric measurement ($OD_{590nm}$) in an ELISA reader. The lowest concentration of the tested extracts required to prevent the formation of ≥50% of the biofilm was considered as minimum biofilm inhibitory concentration ($MBIC_{50}$).[31]

528

## Results and Discussion

*Plants extraction yield*
In this study, the ethanol crude extracts were further fractionated using different polarity solvents including hexane, ethyl acetate and water. Yield of the tested plants extracts, and fractions are shown in Table 1. In general, the yield of hexane fractions was lower than the yields of water and ethyl acetate fractions. The ethyl acetate fractions of *A. officinalis* and *P. guaja* have the highest yield (38.14 and 58.90%, respectively), whereas *T. cordata* water fraction gave the highest yield (52.27%).

The fractionation procedure was performed to obtain fractions containing compounds distributed according to their polarity. Variable polarity solvents were used including hexane, ethyl acetate and water. None polar compositions was extracted using hexane while medium and high polar compositions were extracted using ethyl acetate and water.[32] The high yield in *A. officinalis* and *P. guaja* extract using ethyl acetate and water indicated that the components present in these species are medium to high polar components.

*Antibacterial activity of A. officinalis, T. cordata and P. guaja*
The antibacterial activity was evaluated using disc diffusion method and microdilution method. In general, *P. guaja* exhibited stronger antibacterial activity than *A. officinalis* and *T. cordata* (Table 2). No remarkable antibacterial activity was observed for *A. officinalis* and *T. cordata*. In particular, weak antibacterial activity of *A. officinalis* extract and fractions was reported with maximum inhibition zone of 8.33 mm (Table 2). Similar result was indicated for *T. cordata* extract and fractions with maximum inhibition of 8.67 mm (Table 2). The antibacterial activity of *A. officinalis* ethanol extract and fractions was observed against *P. aeruginosa* (ESBL) and *K. pneumonia* (BL) while *T. cordata* ethanol extract and fractions showed no antibacterial activity (0.0 mm) against all strains tested except *E. coli* (BL).

Our results regarding the antibacterial activity of *A. officinalis* and *T. cordata* are in parallel to that reported previously. Naovi[33] showed that the flower, leaf, root and seed extracts of *A. officinalis* at 10.0 mg/mL exhibited no antibacterial activity against *Corynebacterium diphtheriae, Diplococcus pneumoniae, Staphylococcus aureus, Streptococcus pyogenes* and *Streptococcus viridans*. Ozturk and Ercisli[34] showed that the aqueous extracts from aerial parts of *A. officinalis* had no antibacterial effects, whereas the methanol extracts exhibited moderate antibacterial activity against *Acidovorax facilis, Bacillus sp., Enterobacter hormachei*, and *Kocuria rosea* with maximum inhibition zone of 12 mm.

Regarding the antibacterial activity of *T. cordata*, Fitsiou *et al*.,[35] showed that *T. cordata* essential oil exhibited no antibacterial activity against *Staphylococcus aureus* ATCC 25923, *Sarcina lutea* ATCC 9341, *Bacillus cereus* ATCC 14579, *Escherichia coli* ATCC 25922. At 50 µg/mL, *T. cordata* hydroethanol extract exhibited weak antibacterial activity against methicillin-resistant *Staphylococcus aureus* (MRSA), *Staphylococcus aureus, Escherichia coli, Pseudomonas aeruginosa,* and *Mycobacterium intracellulare* with less than 37% inhibition activity.[36]

*P. guaja* ethanol extract showed stronger antibacterial activity. All strains tested exhibited similar susceptibility with inhibition zone ranged from 9.83 to 11.5 mm. As shown in Table 2, all fractions of *P. guaja* exhibited antibacterial activity against the tested strains except hexane fraction. Ethyl acetate and water fractions appear to possess similar inhibitory effect and all strains were equally susceptible to the tested fractions. Maximum activity of *P. guaja* was observed when ethyl acetate fraction was tested against *E. coli* (BL) with inhibition zone of 13.17 mm. In other study, water, methanol and chloroform extracts of *P. guaja* leaves were reported to inhibit the growth of *Staphylococcus aureus* and beta-streptococcus group A.[37] It was reported that, the hexane, methanol, ethanol, and water extracts of *P. guaja* leaves exhibited antibacterial activity against gram positive bacteria *B. cereus and S. aureus* (mean zones of inhibition of 8.27 and 12.3 mm, and 6.11 and 11.0 mm, respectively) but had no activity against gram negative bacteria.[38]

Minimum Inhibitory Concentration (MIC) of the plant extract against the tested isolates is shown in Table 3. The MIC ranged from 0.80 to 12.50 mg/mL. The MIC values of *A. officinalis* and *T. cordata* extracts and fractions were in the range between 6.25 to 12.50 mg/mL indicating weak antibacterial activity (Table 3). Previous reports showed that *A. officinalis* and *T. cordata* possess bacteriostatic and bactericidal activities.[34,39] Rezaei *et al*.,[39] showed that hydroalcoholic extract of *A. officinalis* possesses bacteriostatic and bactericidal activities against *Staphylococcus aureus* at 330 and 660 µg/mL, respectively, whereas the extract was inactive against Gram negative bacteria including *Listeria sp, Pseudomonas*, and *Escherichia coli*. Ozturk and Ercisli,[34] showed that, the aqueous extracts from aerial parts of *A. officinalis* had no antibacterial effects; whereas, the methanol extracts exhibited significant antibacterial activity against *Acidovorax facilis, Bacillus sp., Enterobacter hormachei,* and *Kocuria rosea* with range of MIC values (62.50 to 500 µg/mL). MIC and MBC of *A. officinalis* ethanol extract against *P. aeruginosa* was found 62.5 mg/L[40]. The ethanol extracts of *T. cordata* produced bacteriostatic activity against gram negative bacteria including *Listeria ivanovii, Serratia rubidaea, Listeria innocua* with range of MIC values (100 to 400 µg/mL) however, against *E. coli, P. aeruginosa* and the gram positive; *Enterococcus raffinosus, Lactobacillus rhamnosus, S. epidermis, Brochothrix thermosphacta* and *Paenobacillus larvae*, the MIC was more than 1000 µg/mL.[41]

Lower MIC values were observed for *P. guaja* (1.56 to 6.25 mg/mL). The MIC values of *P. guaja* extracts and fractions were 1.56 mg/mL against all strains tested. The exception of this is the MIC values against *P. aeruginosa* (3.13 mg/mL) and *E. cloacea* (6.25 mg/mL). Previous reports showed that *P. guaja* extracts possess antibacterial activity with MIC values ranging from 150 µg/mL to 4 mg/mL.[42–44] Ethanol and water extracts of *P. guaja* were reported with bacteriostatic activity against *L. monocytogenes, S. aureus, V. parahaemolyticus,* and *A. faecalis* with MIC values ranged from 0.1 to 0.4 mg/mL and 0.2-0.7 mg/mL, respectively.[45] The variation in MIC values revealed in other reports for the same plant extracts may be due to the variation in their chemical composition, solvents and methods of extractions as well as the pathogenic strains used.[27,46]

*Antibacterial activity of the herbal formula comprising A. officinalis, T. cordata and P. guaja*
The antibacterial activity of the herbal formula comprising *A. officinalis, T. cordata* and *P. guaja* in equal proportion was evaluated using disc diffusion method (Table 2). The antibacterial activity of the water fraction was higher than the antibacterial activity of the ethanol extract. In addition, the antibacterial activity of *A. officinalis, T. cordata* and *P. guaja* extracts and fractions combined in equal proportion increased compared to their activities when they were evaluated individually. The inhibition zones observed for the combined ethanol extracts ranged from 9.33 to 16.17 mm while the zone of inhibition ranged from 0.0 to 11.5 mm when these extracts were evaluated individually. The antibacterial activity for the combined fractions ranged from 9.0 to 19.17 mm.

Maximum inhibitory activity was observed when water fractions were combined generating the range of inhibition zones between 11.83 and 19.17 mm. Combined water fractions was the most active against *E. coli* (BL), *E. coli* (ESBL) and *E. coli* ATCC 25922 leading to inhibition zones of 19.17, 19.0 and 18.67 mm, respectively. *E. cloacae* was the most resistant strain to these combinations with moderate diameter of inhibition zones ranges between 9.0 and 11.83 mm.

**Table 1:** Yield (%) of *A. officinalis*, *T. cordata* and *P. guaja* extracts and fractions

| Solvent | *A. officinalis* | *T. cordata* | *P. guaja* |
|---|---|---|---|
| **EtOH** | 17.20 | 10.56 | 13.04 |
| **Hexane** | 26.51 | 12.12 | 15.95 |
| **Ethyl acetate** | 38.14 | 23.48 | 58.90 |
| **Water** | 32.09 | 52.27 | 22.09 |

**Table 2:** Zone of inhibition (mm) of Plant extracts using disc diffusion method against the tested isolates

| Plant | Solvent of extraction | P. aeruginosa | E. coli (BL) | E. coli (ESBL) | E. coli (ATCC 25922) | K. pneumonia | P. aeruginosa (ATCC 10145) | S. aureus (BL) | E. cloacae |
|---|---|---|---|---|---|---|---|---|---|
| A. officinalis | EtOH | 7.17 ± 0.29 | - | - | - | 7.33 ± 0.58 | - | 7.33 ± 0.58 | - |
| | Hexane | 7.33 ± 0.58 | 8.33 ± 0.29 | - | - | - | 8.17 ± 0.29 | 7.17 ± 0.29 | - |
| | Ethyl acetate | 7.17 ± 0.29 | - | 7.17 ± 0.29 | - | 7.33 ± 0.58 | - | 7.33 ± 0.58 | - |
| | water | 7.17 ± 0.29 | - | - | - | - | 7.33 ± 0.58 | 7.17 ± 0.29 | - |
| T. cordata | EtOH | - | 8.17 ± 0.29 | - | - | - | - | - | - |
| | Hexane | - | 8.17 ± 0.29 | - | - | - | - | - | - |
| | Ethyl acetate | - | 8.67 ± 0.58 | - | - | - | - | - | - |
| | water | - | 7.33 ± 0.58 | 7.17 ± 0.29 | - | - | - | - | - |
| P. guaja | EtOH | 10.67 ± 0.58 | 11.17 ± 0.76 | 11.5 ± 0.50 | 11.0 ± 0.50 | 10.33 ± 0.58 | 11.0 ± 0.50 | 10.67 ± 0.29 | 9.83 ± 0.76 |
| | Hexane | - | - | - | - | - | - | - | - |
| | Ethyl acetate | 12.33 ± 0.58 | 13.17 ± 0.29 | 12.83 ± 0.76 | 12.33 ± 0.29 | 11.83 ± 0.29 | 12.50 ± 0.50 | 11.83 ± 0.29 | 11.33 ± 0.29 |
| | water | 11.33 ± 0.58 | 12.33 ± 0.29 | 11.33 ± 0.58 | 11.83 ± 1.04 | 10.50 ± 0.87 | 11.17 ± 1.26 | 11.00 ± 0.0 | 10.50 ± 0.50 |
| Combination (1:1:1) | EtOH | 16.17 ± 0.29 | 15.50 ± 0.87 | 15.83 ± 0.29 | 14.83 ± 0.29 | 14.67 ± 0.29 | 16.17 ± 0.76 | 15.17 ± 0.29 | 9.33 ± 0.58 |
| | Hexane | 10.67 ± 0.58 | 11.17 ± 0.29 | 11.17 ± 0.29 | 10.83 ± 1.04 | 10.50 ± 0.87 | 10.83 ± 0.76 | 10.67 ± 0.29 | 10.00 ± 0.0 |
| | Ethyl acetate | 15.50 ± 0.50 | 16.33 ± 0.58 | 17.50 ± 0.87 | 17.50 ± 0.5 | 15.67 ± 0.76 | 15.33 ± 0.58 | 14.17 ± 0.29 | 9.00 ± 0.0 |
| | water | 17.67 ± 0.58 | 19.17 ± 0.29 | 19.00 ± 0.50 | 18.67 ± 1.15 | 17.17 ± 1.26 | 17.83 ± 0.29 | 16.17 ± 1.04 | 11.83 ± 0.29 |
| Standard antibiotics | Ciprofloxacin (5 µg) | 24.8 ± 0.8 | 24.5 ± 0.5 | 24.5 ± 0.5 | 22.2 ± 0.3 | 20.5 ± 0.5 | 24.5 ± 0.5 | 22.3 ± 0.6 | 17.8 ± 0.8 |

The MIC values of the combined ethanol extracts and water fractions decreased compared to the MIC values of these extracts and fractions evaluated individually (Table 3). Comparing the MIC values of the *P. guaja*, lower MIC values were observed for the combined ethanol extracts and water fractions against *E. coli* (BL), *E. coli* (ESBL), *E. coli* ATCC 25922, *K. pneumonia* and *S. aureus* (BL) (1.56 to 0.8 mg/mL).

The MIC values of hexane and ethyl acetate fractions against these strains were similar to the MIC values of the *P. guaja* (1.56 mg/mL). Decreased in the MIC values was also observed when the combined extracts and fractions were tested against *P. aeruginosa* (BL) from 3.13 to 6.25 mg/mL. *P aeruginosa* ATCC 10145 and *E. cloacea* showed similar susceptibility to *P. guaja* combined extracts and fractions with 1.56 and 6.25 mg/mL, respectively.

As proposed previously, plant extract with MIC value less than 1.0 mg/mL should be considered as a strong antibacterial agent.[47,48] Accordingly, only the combined ethanol extracts and water fractions can be considered as potent antibacterial agents since the MIC values against *E. coli* (BL), *E. coli* (ESBL), *E. coli* ATCC 25922 and *S. aureus* (BL) were 0.8 mg/mL (each). Interestingly, the combined extracts and fractions showed stronger antibacterial activity. However, this activity is closer to the activity of the *P. guaja* extracts and fractions suggesting that *P. guaja* is the major active components in this formula. The sub-MIC values for the combined extracts and fractions reported in this study indicated that combining two or more extracts may increase the antibacterial activity. The reason might be related to the synergistic interaction between the formula components.[49] In fact, traditional healers use combination between plants to increase the therapeutic effects of a single plant.[50,51]

*Antibiofilm activity of A. officinalis, T. cordata and P. guaja extracts and fractions*

The antibiofilm activity of ethanol extracts and fractions individually and in combination was evaluated using crystal violet assay at sub MIC concentrations (0.2, 0.4, 0.8 mg/mL) against *P. aeruginosa* (Figure 1) and *E. coli* (Figure 2). The results showed dose-dependent inhibition activity for all extracts and fractions tested. At the highest concentration tested (0.8 mg/mL), hexane fraction of *A. officinalis*, ethanol and water fractions *of T. cordata and P. guaja* completely inhibited the formation of *P. aeruginosa* biofilm (>90%). Hexane, ethyl acetate and water fractions of *P. guaja* caused complete inhibition of *E. coli* biofilm. Regarding the combined extracts and fractions, all combined extracts and fractions completely inhibited the formation of *P. aeruginosa* and *E. coli*.

Minimum Biofilm Inhibitory Concentration (MBIC$_{50}$) for the tested extracts and fractions and in combination was determined (Table 4). The lowest MBIC$_{50}$ reported was 0.2 mg/mL for *P. guaja* and the herbal formula extracts and fractions indicating potent antibiofilm activity. Higher MBIC$_{50}$ was observed for *T. cordata* and *A. officinalis* extracts and fractions (0.4 mg/mL, each). In another study, *A. officinalis* extracts at concentrations equal to MIC (62.5 mg/L) or higher inhibited *P. aeruginosa* biofilm by 87.3 and 93.7%, respectively.[40] There were no available data about the antibiofilm activity of *T. cordata* and *P. guaja* and the formula.

**Table 3:** Minimum Inhibitory Concentration (MIC) (mg/mL) of the plant extract against the tested isolates

| Plant species | Solvent of extraction | P. aeruginosa | E. coli BL | E. coli ESBL | E. coli (ATCC 25922) | K. pneumonia | P. aeruginosa (ATCC 10145) | S. aureus (BL) | E. cloacae |
|---|---|---|---|---|---|---|---|---|---|
| **A. officinalis** | EtOH | 6.25 | 6.25 | 6.25 | 12.5 | 6.25 | 6.25 | 6.25 | 12.5 |
| | Hexane | 6.25 | 6.25 | 6.25 | 6.25 | 6.25 | 6.25 | 6.25 | 12.5 |
| | Ethyl acetate | 6.25 | 6.25 | 6.25 | 6.25 | 6.25 | 6.25 | 6.25 | 12.5 |
| | water | 6.25 | 6.25 | 6.25 | 6.25 | 6.25 | 6.25 | 6.25 | 12.5 |
| **T. cordata** | EtOH | 6.25 | 12.5 | 12.5 | 12.5 | 12.5 | 12.5 | 12.5 | 12.5 |
| | Hexane | 6.25 | 12.5 | 12.5 | 12.5 | 12.5 | 12.5 | 6.25 | 12.5 |
| | Ethyl acetate | 6.25 | 12.5 | 12.5 | 12.5 | 12.5 | 12.5 | 6.25 | 12.5 |
| | water | 12.5 | 12.5 | 12.5 | 12.5 | 12.5 | 12.5 | 12.5 | 12.5 |
| **P. guaja** | EtOH | 3.13 | 1.56 | 1.56 | 1.56 | 1.56 | 1.56 | 1.56 | 6.25 |
| | Hexane | 3.13 | 1.56 | 1.56 | 1.56 | 1.56 | 1.56 | 1.56 | 6.25 |
| | Ethyl acetate | 3.13 | 1.56 | 1.56 | 1.56 | 1.56 | 1.56 | 1.56 | 6.25 |
| | water | 3.13 | 1.56 | 1.56 | 1.56 | 1.56 | 1.56 | 1.56 | 6.25 |
| **Combination (1:1:1)** | EtOH | 1.56 | 0.8 | 0.8 | 0.8 | 0.8 | 1.56 | 0.8 | 6.25 |
| | Hexane | 1.56 | 1.56 | 1.56 | 1.56 | 1.56 | 1.56 | 1.56 | 6.25 |
| | Ethyl acetate | 1.56 | 1.56 | 1.56 | 1.56 | 1.56 | 1.56 | 1.56 | 6.25 |
| | water | 1.56 | 0.8 | 0.8 | 0.8 | 0.8 | 1.56 | 0.8 | 6.25 |

**Table 4:** Minimum Biofilm Inhibitory Concentration (MBIC$_{50}$) of the plant extract

| Plant species | Solvent of extraction | P. aeruginosa | E. coli |
|---|---|---|---|
| A. officinalis | EtOH | 0.4 | 0.4 |
| | Hexane | 0.4 | 0.4 |
| | Ethyl acetate | 0.4 | 0.4 |
| | water | 0.2 | 0.4 |
| T. cordata | EtOH | 0.4 | 0.4 |
| | Hexane | 0.4 | 0.4 |
| | Ethyl acetate | 0.4 | 0.4 |
| | water | 0.4 | 0.4 |
| P. guaja | EtOH | 0.2 | 0.2 |
| | Hexane | 0.2 | 0.2 |
| | Ethyl acetate | 0.2 | 0.2 |
| | water | 0.2 | 0.2 |
| Combination (1:1:1) | EtOH | 0.2 | 0.2 |
| | Hexane | 0.2 | 0.2 |
| | Ethyl acetate | 0.2 | 0.2 |
| | water | 0.2 | 0.2 |

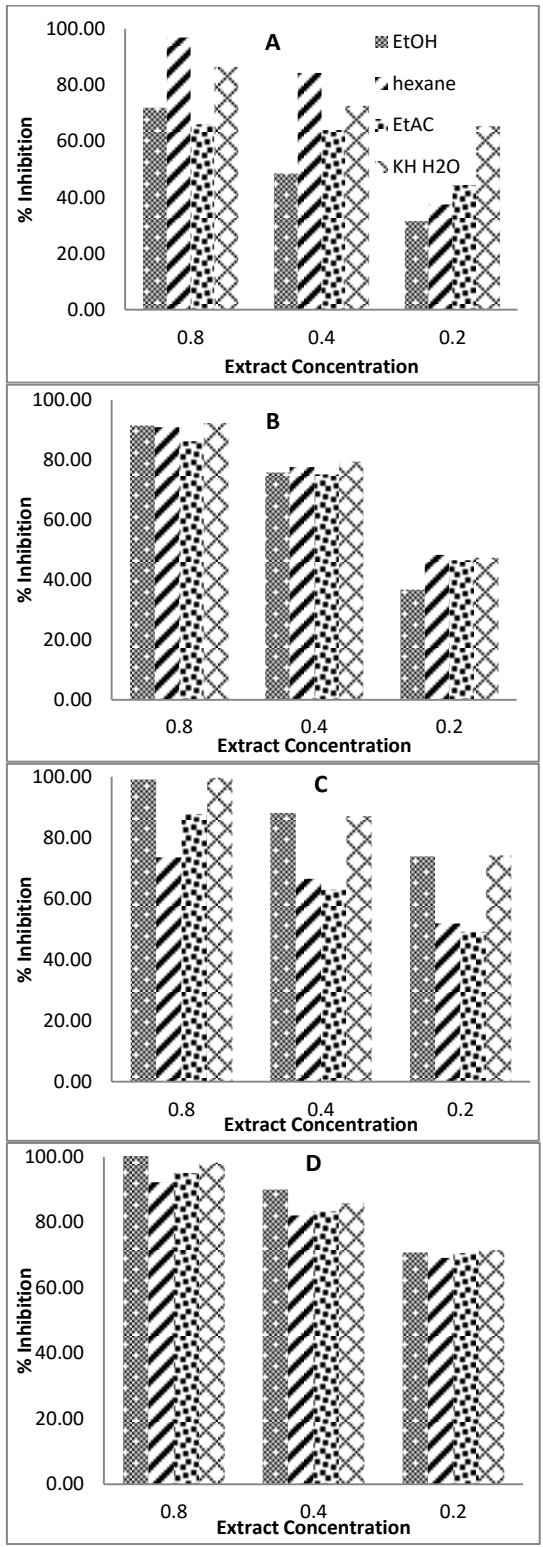
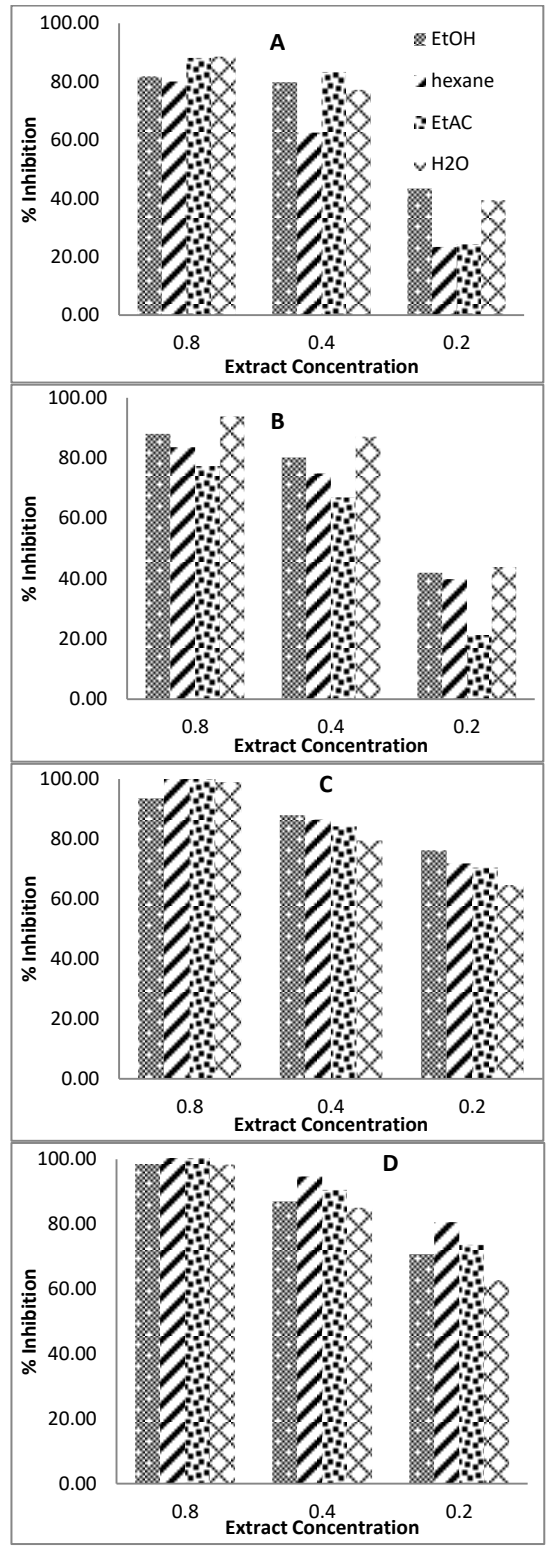

**Figure 1:** Percent of *P. aeruginosa* biofilm inhibition, A: *A. officinalis,* B: *T. cordata,* C: *P. guaja;* D: combination (1:1:1)

**Figure 2:** Percent of *E. coli* biofilm inhibition, A: *A. officinalis,* B: *T. cordata,* C: *P. guaja;* D: combination (1:1:1)

**Conclusion**

Plants selected in this study have been used traditionally for the treatment of respiratory infections. *A. officinalis* flowers and *T. cordata* leaves extracts exhibited weak antibacterial activity while *P. guaja* exhibited moderate antibacterial activity against respiratory infection causing bacteria. Combining these plants extracts in equal proportion provided stronger antibacterial and antibiofilm activities. Therefore, this study provides a valuable scientific knowledge to support the use of plants in combination rather than individually. Further investigations are required to evaluate the cytotoxicity of this combination and to include other pathogenic bacteria and viruses.

**Conflict of interest**

The authors declare no conflict of interest.

**Authors' Declaration**

The authors hereby declare that the work presented in this article is original and that any liability for claims relating to the content of this article will be borne by them.

**References**

1. Ferkol T and Schraufnagel D. The global burden of respiratory disease. Ann Am Thorac Soc. 2014; 11(3):404–406.
2. Mogyoros M. Challenges of managed care organizations in treating respiratory tract infections in an age of antibiotic resistance. Am J Manag Care. 2001; 7(6 Suppl):S163-169.
3. Rudan I, Boschi-Pinto C, Biloglav Z, Mulholland K, Campbell H. Epidemiology and etiology of childhood pneumonia. Bull World Health Org. 2008; 86:408-416B.
4. Rudan I, O'brien KL, Nair H, Liu L, Theodoratou E, Qazi S, Lukšić I, Walker CL, Black RE, Campbell H. Epidemiology and etiology of childhood pneumonia in 2010: estimates of incidence, severe morbidity, mortality, underlying risk factors and causative pathogens for 192 countries. J Glob Health. 2013;3(1): 010401
5. Dasaraju PV and Liu C. Infections of the respiratory system. Med Microbiol. 1996; 4. Dasaraju PV, Liu C. Infections of the Respiratory System. In: Baron S, editor. Medical Microbiology. 4th edition. Galveston (TX): University of Texas Medical Branch at Galveston; 1996. Chapter 93
6. WHO. Global priority list of antibiotic-resistant bacteria to guide research, discovery, and development of new antibiotics. 2017.
7. Al-Asoufi A, Khlaifat A, Al Tarawneh A, Alsharafa K, Al-Limoun M, Khleifat K. Bacterial Quality of Urinary Tract Infections in Diabetic and Non Diabetics of the Population of Ma'an Province, Jordan. Pak J Biol Sci. 2017; 20(4):179–188.
8. Jarab AS, Mukattash TL, Nusairat B, Shawaqfeh M, Farha RA. Patterns of antibiotic use and administration in hospitalized patients in Jordan. Saudi Pharm J. 2018; 26(6):764–770.
9. Khleifat K, Homady MH, Tarawneh KA, Shakhanbeh J. Effect of *Ferula hormonis* extract on social aggression, fertility and some physiological parameters in prepubertal male mice. Endocr J. 2001; 48(4):473–482.
10. Homady MH, Khleifat KM, Tarawneh KA, Al-Raheil IA. Reproductive toxicity and infertility effect of *Ferula hormonis* extracts in mice. Theriogenol. 2002; 57(9):2247–2256.
11. Shaikh BT and Hatcher J. Complementary and alternative medicine in Pakistan: prospects and limitations. Evid-Based Compl Altern Med. 2005; 2(2):139–142.
12. Issa RA and Basheti IA. Herbal Medicine Use by People in Jordan: Exploring Believes and Knowledge of Herbalists and Their Customers. J Biol Sci. 2017; 17(8):400–409.
13. Abdelhalim A, Aburjai T, Hanrahan J, Abdel-Halim H. Medicinal plants used by traditional healers in Jordan, the Tafila region. Pharmacogn Mag. 2017; 13(Suppl 1):S95.
14. Parasuraman S, Thing GS, Dhanaraj SA. Polyherbal formulation: Concept of ayurveda. Pharmacogn Rev. 2014; 8(16):73.
15. Heinrich M. Medicinal Plants of the World. Vol. 2: Chemical Constituents, Traditional and Modern Medicinal Uses, Ivan A. Ross, Humana Press, Totowa, NJ, USA, 2001. ISBN 0-89603-877-7; US $99.50 (hbk), 487 pp.; cross-reference of names, index, glossary; some colour, a. J Ethnopharmacol. 2001; 3(76):309.
16. Shah SMA, Akhtar N, Akram M, Shah PA, Saeed T, Ahmed K, Asif HM. Pharmacological activity of *Althaea officinalis* L. J Med Plants Res. 2011; 5(24):5662–5666.
17. Blumenthal M, Goldberg A, Brinckmann J. Hawthorn Herb Med Expand. Comm. E Monogr Am Bot Counc Integr Med Commun. Newton, MA. 2000. 182–189 p.
18. Al-Essa MK, Mohammed FI, Shafagoj YA, Afifi FU. Studies on the Direct Effects of the Alcohol Extract of *Tilia cordata*. on Dispersed Intestinal Smooth Muscle Cells of Guinea Pig. Pharm Biol. 2007;45(3):246–250.
19. Jansen PCM and Mendes O. Plantas medicinais seu uso tradicional em Mocambique. Unknown Publisher; 1990: 115
20. Smith NJH, Williams JT, Plucknett DL, Talbot JP. Tropical forests and their crops. Cornell University Press; 2018: 8
21. Gutiérrez RMP, Mitchell S, Solis RV. *Psidium guajava*: a review of its traditional uses, phytochemistry and pharmacology. J Ethnopharmacol. 2008; 117(1):1–27.
22. Li J, Chen F, Luo J. GC-MS analysis of essential oil from the leaves of *Psidium guajava*. Zhong yao cai Zhongyaocai J. Chinese Med. Mater. 1999;22(2):78.
23. Murti Y and Sharma S. Flavonoid: A Pharmacologically Significant Scaffold. World Journal of Pharmacy and Pharmaceutical Sciences, 2017;6(5): 488-504.
24. Anand V, Kumar V, Kumar S, Hedina A. Phytopharmacological overview of Psidium guajava Linn. Pharmacogn J. 2016; 8(4): 314-320
25. Bibi Y, Nisa S, Chaudhary FM, Zia M. Antibacterial activity of some selected medicinal plants of Pakistan. BMC Compl Altern Med. 2011; 11(1):52.
26. Qaralleh H, Idid S, Saad S, Susanti D, Taher M, Khleifat K. Antifungal and Antibacterial Activities of Four Malaysian Sponge Species (Petrosiidae). J Mycol Med. 2010; 20(4).
27. Qaralleh HN. Chemical composition and antibacterial activity of Origanum ramonense essential oil on the β-lactamase and extended-spectrum β-lactamase urinary tract isolates. Bangladesh J Pharmacol. 2018; 13(3):280–286.
28. Althunibat OY, Qaralleh Q, Al-Dalin SYA, Abboud M, Khleifat K, Majali IS, Aldal'in HK, Rayyan WA, Jaafraa A. Effect of Thymol and Carvacrol, the Major Components of *Thymus capitatus* on the Growth of *Pseudomonas aeruginosa*. J Pure Appl Microbiol. 2016; 10(1):367–374.
29. Qaralleh HN, Abboud MM, Khleifat KM, Tarawneh KA, Althunibat OY. Antibacterial activity in vitro of *Thymus capitatus* from Jordan. Pak J Pharm Sci. 2009; 22(3).
30. Hossain MA, Lee S-J, Park NH, Mechesso AF, Birhanu BT, Kang J, Reza MA, Suh JW, Park SC. Impact of phenolic compounds in the acyl homoserine lactone-mediated quorum sensing regulatory pathways. Sci Rep. 2017; 7(1):10618.
31. Lopes LAA, dos Santos Rodrigues JB, Magnani M, de Souza EL, de Siqueira-Júnior JP. Inhibitory effects of flavonoids on biofilm formation by *Staphylococcus aureus* that overexpresses efflux protein genes. Microb Pathog. 2017; 107:193–197.

32. Nwabueze TU and Okocha KS. Extraction performances of polar and non-polar solvents on the physical and chemical indices of African breadfruit (*Treculia africana*) seed oil. Afr J Food Sci. 2008; 2(10):119–125.
33. Naqvi S, Khan MSY, Vohora SB. Anti-bacterial, anti-fungal and anthelmintic investigations on Indian medicinal plants. Fitoterapia. 1991; 62:221–228.
34. Ozturk S and Ercisli S. Antibacterial Activity of Aqueous and Methanol Extracts of *Althaea officinalis* and *Althaea cannabina* from Turkey. Pharm Biol. 2007; 45(3):235–240.
35. Fitsiou Ioanna, Tzakou O, Hancianu M, Poiata A. Volatile constituents and antimicrobial activity of Tilia tomentosa Moench and Tilia cordata Miller oils. J Essent Oil Res. 2007; 19(2):183–185.
36. Gawad SMA, Hetta MH, Ross SA, Badria FA. Antiprotozoal and antimicrobial activity of selected medicinal plants growing in upper egypt, beni-suef region. World J Pharm Pharm Sci. 2015; 4:1720–1740.
37. Jaiarj P, Khoohaswan P, Wongkrajang Y, Peungvicha P, Suriyawong P, Saraya MS, Ruangsomboon O. Anticough and antimicrobial activities of *Psidium guajava* Linn. leaf extract. J Ethnopharmacol. 1999; 67(2):203–212.
38. Biswas B, Rogers K, McLaughlin F, Daniels D, Yadav A. Antimicrobial activities of leaf extracts of guava (*Psidium guajava* L.) on two gram-negative and gram-positive bacteria. Int J Microbiol. 2013;2013.
39. Rezaei M, Dadgar Z, Noori-Zadeh A, Mesbah-Namin SA, Pakzad I, Davodian E. Evaluation of the antibacterial activity of the *Althaea officinalis* L. leaf extract and its wound healing potency in the rat model of excision wound creation. Avicenna J Phytomed. 2015; 5(2):105.
40. Aminnezhad S, Kermanshahi RK, Ranjbar R. Effect of *Althaea officinalis* extract on growth and biofilm formation in *Pseudomonas aeruginosa*. J Pure Appl Microbiol. 2016; 10(3):1857–1864.
41. Vatlák A, Kolesárová A, Vukovic N, Rovná K, Petrová J, Vimmerová V, Hleba L, Mellen M, Kačániová M. Antimicrobial activity of medicinal plants against different strains of bacteria. J Microbiol Biotechnol Food Sci. 2014; 3:174.
42. Prabu GR, Gnanamani A, Sadulla S. Guaijaverin–a plant flavonoid as potential antiplaque agent against *Streptococcus mutans*. J Appl Microbiol. 2006; 101(2):487–495.
43. Arima H and Danno G. Isolation of antimicrobial compounds from guava (*Psidium guajava* L.) and their structural elucidation. Biosci Biotechnol Biochem. 2002; 66(8):1727–1730.
44. Sanches NR, Garcia Cortez DA, Schiavini MS, Nakamura CV, Dias Filho BP. An evaluation of antibacterial activities of *Psidium guajava* (L.). Braz Arch Biol Technol. 2005; 48(3):429–436.
45. Mahfuzul Hoque MD, Bari ML, Inatsu Y, Juneja VK, Kawamoto S. Antibacterial activity of guava (*Psidium guajava* L.) and neem (*Azadirachta indica* A. Juss.) extracts against foodborne pathogens and spoilage bacteria. Foodborne Pathog Dis. 2007; 4(4):481–488.
46. Hussain AI, Anwar F, Nigam PS, Ashraf M, Gilani AH. Seasonal variation in content, chemical composition and antimicrobial and cytotoxic activities of essential oils from four Mentha species. J Sci Food Agric. 2010; 90(11):1827–1836.
47. Rios JL and Recio MC. Medicinal plants and antimicrobial activity. J Ethnopharmacol. 2005; 100(1–2):80–84.
48. Van Vuuren SF. Antimicrobial activity of South African medicinal plants. J Ethnopharmacol. 2008; 119(3):462–472.
49. Chusri S, Tongrod S, Saising J, Mordmuang A, Limsuwan S, Sanpinit S, Voravuthikunchai SP. Antibacterial and anti-biofilm effects of a polyherbal formula and its constituents against coagulase-negative and-positive staphylococci isolated from bovine mastitis. J Appl Anim Res. 2017; 45(1):364–372.
50. Simbo DJ. An ethnobotanical survey of medicinal plants in Babungo, Northwest Region, Cameroon. J Ethnobiol Ethnomed. 2010; 6(1):8.
51. Neamsuvan O, Tuwaemaengae T, Bensulong F, Asae A, Mosamae K. A survey of folk remedies for gastrointestinal tract diseases from Thailand's three southern border provinces. J Ethnopharmacol. 2012; 144(1):11–21.